\documentstyle[aps]{revtex}
\input epsf
\input psfig.sty

\begin{document}
\draft

\title{FERROMAGNETIC LIQUID THIN FILMS UNDER APPLIED FIELD\\}
\author{S. Banerjee and M. Widom}
\address{Department of Physics, Carnegie Mellon University,
Pittsburgh, Pennsylvania 15213}

\date{\today}
\maketitle
\begin{abstract}
Theoretical calculations, computer simulations and experiments
indicate the possible existence of a ferromagnetic liquid state,
although definitive experimental evidence is lacking. Should such a
state exist, demagnetization effects would force a nontrivial magnetization
texture. Since liquid droplets are deformable, the droplet shape is
coupled with the magnetization texture. In a thin-film
geometry in zero applied field, the droplet has a circular shape and a 
rotating magnetization texture with a point vortex at the center. We 
calculate the elongation and magnetization texture of such ferromagnetic 
thin film liquid droplet confined between two parallel plates under a 
weak applied magnetic field. The vortex stretches into a domain wall and 
exchange forces break the reflection symmetry. This behavior contrasts 
qualitatively and quantitatively with the elongation of paramagnetic 
thin films.
\end{abstract}

\section{introduction}
\label{intro}

The study of a possible ferromagnetic liquid state is a problem of 
considerable interest. In such a spontaneously magnetized
liquid state, long range magnetic order would exist in the liquid without
application of any external field. The existence of such a liquid
state has been indicated by mean field 
calculations~\cite{sano,tsebers,widom,dietrich,groh} and computer simulations 
on strongly polar fluids~\cite{patey,weis1,weis2,stevens}.

Experiments to observe ferromagnetism in liquids with strong
magnetic interactions such as ferrofluids~\cite{rr} have failed so far because
the liquids freeze~\cite{luo,ederbeck} or phase separate~\cite{luo2} well above 
the predicted low temperatures for the onset of 
spontaneous magnetization. Experiments on super-cooled Co-Pd 
alloys~\cite{platzek} do indicate the possibility of 
ferromagnetism in liquids. In this case it is the strong exchange 
interaction and not the dipole interaction that would cause the 
spontaneous magnetization. The experimental evidence regarding Co-Pd is
still not conclusive. 

Although the existence of a ferromagnetic liquid state is yet to be confirmed 
experimentally, spontaneous polarization coupled with other order parameters
has already been observed. Some electrically polarized liquid 
crystals~\cite{liquid-crystal} show a helical ordering of the dipole 
moments in the liquid. In superfluid $^3$He the magnetic moment couples 
to the superconducting order parameter~\cite{helium}. Many superfluid $^3$He 
phases are therefore also magnetically ordered. 

It is interesting to consider the magnetization texture (spatial variation
of the orientation of magnetization) inside a droplet of such a
ferromagnetic liquid~\cite{degennes}. The magnetization texture likes
to avoid poles~\cite{wfb} to minimize its energy. However, this
leads to defects inside the texture. For example, a rotating
magnetization texture with cylindrical symmetry inside a sphere avoids
all poles but has a vortex line running through the center. Near the
vortex of such a texture the magnetization is topologically
unstable~\cite{mermin} and might escape into the third
dimension~\cite{hubert} with a nonzero component along the vortex
line. Whether this happens depends on the balance between demagnetizing
and vortex energies. Simulated annealing of the magnetization inside a
cubic box suggests that replacing vortices with point defects may be
favorable~\cite{groh}. Any defect is likely to have a system-shape
dependent energy cost causing a deformable liquid droplet
to deviate from a spherical shape. The complete calculation of the
shape of an unconfined ferromagnetic liquid droplet in three-dimensions
coupled with the calculation of its magnetization texture remains an
unsolved problem.

This problem has a simple solution in two-dimensions in zero field. The 
magnetization texture inside any soft (zero anisotropy) ferromagnetic solid 
thin film is given by van den Berg's algorithm~\cite{vandenberg} which avoids 
all poles, and thus all magnetostatic energy, at the expense of a domain wall 
through the film. A liquid droplet, which 
can change its shape, prefers a circular shape to minimize its surface energy. 
The magnetization lines inside a circle form concentric circles according
to van den Berg's algorithm. For a circular shape the domain wall energy is 
also minimized because the domain wall shrinks to a point vortex. The circle
thus solves the coupled texture and shape problem in zero field.

Our goal is to analyze the shape of a ferromagnetic liquid droplet under
weak applied fields. Since the texture and shape of a three-dimensional drop is
not precisely known at present, we concentrate on the two-dimensional case of a 
droplet confined between two parallel plates with spacing much smaller than the
droplet  diameter. When a field is applied parallel to the plates the
magnetization texture distorts to exclude the magnetic field from the bulk of
the droplet. Bryant and Suhl~\cite{suhl} calculated the texture of solid 
circular and elliptic thin films under applied field. We adapt their result 
to ferromagnetic liquids by letting the shape vary as the field is applied. 
We find that the droplet elongates to reduce its magnetostatic energy.
The equilibrium shape is reached when the magnetostatic energy plus the surface 
energy is minimized.

In Sec.~\ref{calculation} we calculate elongation of a ferromagnetic 
thin film as a function of its undeformed radius, thickness, the 
saturation magnetization and the applied field. In Sec.~\ref{contrast} 
we contrast this behavior with that of paramagnetic thin films under 
applied fields and discuss why this contrasting behavior occurs. In 
Sec.~\ref{symbreak} we discuss
the symmetry breaking of the droplet shape under exchange forces and its
dependence on the applied field. Finally, Sec.~\ref{conclusion} summarizes 
our results.

\section{Calculation}
\label{calculation}

Consider a homogeneous model liquid such that every point in the liquid 
has a saturation magnetization $M_s$ with no anisotropy. Let a droplet of
such a liquid in a spontaneously magnetized state be confined between 
two parallel plates as shown in Fig. 1(a). The thickness spacing $\Delta$ 
between the plates is kept much smaller than the diameter of the undeformed 
droplet so that the droplet has a very small aspect ratio 
$p \equiv \Delta /2r_0$. Figure 1(b) shows a droplet elongating under 
applied field. To a good approximation the droplet shape is an ellipse 
for small elongation. We define the elongation as
\begin{equation}
\epsilon \equiv a/b -1,
\end{equation}
where $a$ and $b$ are the semimajor and semiminor axes of the ellipse. 
We choose a coordinate system with the field in the $\hat x$ direction 
and $\hat z$ normal to the plates. The origin is at the center of the 
droplet. 

Under zero  external field  this thin film droplet has a circular shape. The 
magnetization texture ${\bf M}({\bf r})$, given by van den Berg's 
algorithm~\cite{vandenberg}, 
forms circles concentric  with the droplet boundary. Because this texture is 
divergence-free and everywhere tangent to the surface, there are no magnetic
poles~\cite{wfb} and hence no demagnetizing field. This texture therefore 
achieves the lowest possible magnetostatic energy.
When an external field is applied, the circular magnetization texture 
distorts to
reach a new equilibrium configuration. The new texture exhibits poles, making 
the magnetostatic energy shape dependent. The droplet elongates to reduce this 
shape dependent magnetostatic energy and an equilibrium shape is reached when the 
magnetostatic plus surface energy is minimized. We neglect magnetic exchange
energy because it is small relative to the magnetostatic energy as discussed in
Sec.~\ref{symbreak}. Our goal is to 
calculate this elongation as a function of the applied field and the various 
parameters for the droplet.

To calculate the magnetostatic energy of the droplet we make use of the 
calculations by Bryant and Suhl~\cite{suhl} for the magnetization texture 
of elliptic solid ferromagnetic thin films under applied field. Their 
calculations for the magnetization texture are for fixed shape. We let 
the shape of the thin film droplet vary and calculate its energy as a 
function of the shape for small elongation. The elongation is then 
calculated by minimizing the total energy with respect to elongation.

Bryant and Suhl  calculate the equilibrium magnetization texture 
of an elliptical ferromagnetic thin film in applied field by drawing 
analogy to the case of conductors in applied electric field. A 
ferromagnet is equivalent to a conductor because it has effectively 
an infinite permeability. When a conducting film is placed in an 
applied electric field, it expels the electric field by creating an 
induced charge density on its surface. A soft ferromagnet will 
similarly expel magnetic field if a magnetization texture can be 
found exhibiting the necessary surface pole density. These charges 
$\sigma_M(x,y)$ arise from the components of ${\bf M}$ normal to the top 
and bottom surfaces of the droplet and there are no poles within the 
bulk of the droplet. However, for a thin film we treat the charge as 
if it permeates through the film and the magnetization texture is 
restricted to the $x-y$ plane so that $M_z=0$. The magnetization 
texture then satisfies the following pseudo-divergence 
equation:
\begin{equation}
\label{divergence}
\nabla \cdot {\bf M}\equiv {\partial M_x \over \partial x}+{\partial M_y 
\over \partial y} =-\rho_M=-\sigma_M(x,y) / \Delta.
\end{equation}
The above equation is exact in the limit of zero thickness. It remains 
a good approximation provided the the film has a small aspect 
ratio  ($\Delta /2r_0<<1$), because strong demagnetizing fields prevent 
the magnetization from tilting far out of the $x-y$ plane. The partial 
differential Eq.~(\ref{divergence}) can  be integrated, in principle, 
to find a solution for ${\bf M}({\bf r})$ if it exists.

For an elliptic thin film, the charge density that expels the field can 
be written in a convenient form as
\begin{equation}
\label{sigma}
\sigma_M(x,y)={4 M_s \Delta E_x x \over \pi a^2 {\big ( }1-({x/a})^2-({y/b})^2
{\big )}^{1/2}},
\end{equation}
where 
\begin{equation}
\label{field}
E_x=H_0 {\biggl (} {a^2-b^2 \over 8 M_s b \Delta ( {\tilde K}-{\tilde E})} 
{\biggl )}
\end{equation}
is the reduced applied field. ${\tilde K}$ and ${\tilde E}$ are complete
elliptic integrals of the first and second kind, respectively, of the argument
$(1-b^2/a^2)^{1/2}$.  For $E_x \le 1$, Eq.~(\ref{divergence}) can be 
integrated to find the magnetization texture which expels the magnetic field.
A C-shaped domain wall (see Fig. 2) appears inside the droplet for nonzero 
field. For $E_x >1$, the domain wall intersects the boundary of the film and
the field penetrates the interior of the film.

We use this result to calculate the magnetostatic energy of a liquid droplet. Since
the above calculations are for zero anisotropy, they also apply to liquids.
We begin with the general expression for the total magnetostatic energy of any 
magnetization distribution under an applied field 
\begin{equation}
\label{mag-nrg}
E_M=-\int d\tau~ {\bf H}_0 \cdot {\bf M} - {1 \over 2} \int d\tau~ {\bf H}_D 
\cdot {\bf M}.
\end{equation}
The first term on the right hand side of~(\ref{mag-nrg}) is the energy of 
the applied field ${\bf H}_0$ acting on magnetization ${\bf M}$. The 
second term is the self-energy (hence the factor of 1/2) due to the 
magnetization interacting with its own demagnetizing field ${\bf H}_D$. 
For $E_x \le 1$, ${\bf H}_D=-{\bf H}_0$ because the magnetization 
expels the field, therefore the magnetostatic energy
\begin{equation}
\label{mag-nrg2}
E_M=- {1 \over 2} \int d\tau~ {\bf H}_0 \cdot {\bf M}.
\end{equation}
Using $\rho_M=-\nabla \cdot {\bf M}$, writing ${\bf H}_0=-\nabla \phi_0$, and 
integrating by parts, the magnetostatic energy in~(\ref{mag-nrg2}) can be 
rewritten
\begin{equation}
\label{mag-nrg3}
E_M= {1 \over 2} \int d\tau~\phi_0~\rho_M,
\end{equation}
where $\phi_0=-{\bf H}_0 x$ is the potential due to ${\bf H}_0$. Using 
$\rho_M=\sigma_M/\Delta$ and Eqs.~(\ref{sigma}) and (\ref{field}) 
and calculating the integral in~(\ref{mag-nrg3}) gives
\begin{equation}
\label{mag-nrg3.5}
E_M=-{32 \over 3} E_x^2 M_s^2 \Delta^2 a b^2 {{\tilde K}-{\tilde E}  \over a^2-b^2} 
= -{1 \over 6} H_0^2 a {a^2-b^2 \over {\tilde K}-{\tilde E}},
\end{equation}

Assume that the applied field is small so that $\epsilon << 1$. Expanding
the result in~(\ref{mag-nrg3.5}) for small $\epsilon$ gives
\begin{equation}
\label{mag-nrg4}
E_M = - {2 H_0^2 r_0^3 \over 3 \pi} (1 + {3\over 4} \epsilon),
\end{equation}
with corrections of higher order in $\epsilon$ and 
higher order in the aspect ratio 
$\Delta /2r_0$. As expected, the leading  correction to the energy is 
negative, so the droplet elongates to reduce its magnetostatic energy. 
Interestingly, the magnetostatic energy is independent of the saturation 
magnetization $M_s$. This is so because, as long as $E_x \le 1$, the charge 
density $\rho_M$ does not depend on $M_s$. The factor 
of $M_s$ in the numerator of $\sigma_M(x,y)$ in~(\ref{sigma}) cancels 
against the factor of $M_s$ in the denominator of $E_x$ in 
Eq.~(\ref{field}). Physically, the charge density of a given 
distribution is proportional to $M_s$, but the distortion 
that creates $\rho_M$ varies inversely with $M_s$.

The magnetostatic energy is also independent of $\Delta$ to the lowest order in 
the aspect ratio $\Delta /2r_0$. To understand this, note that the expression 
for the reduced applied field $E_x$ has a factor $\Delta$ in the denominator 
implying that thicker films can expel higher values of applied field for a 
given distortion. Thus a thicker film has
a smaller distortion in its texture, for a particular value of applied field,
than a thinner film. This is confirmed by the factor of the $\Delta$ in the
denominator of the charge density $\rho_M$. When the energy density is 
integrated over the volume it yields a magnetostatic energy independent of 
$\Delta$.

As the droplet elongates, its perimeter increases and is given by
\begin{equation}
S= 2 \pi r_0  (1 + {3\over 16} \epsilon^2),
\end{equation}
with higher order corrections in $\epsilon$. The variation of the 
perimeter is quadratic in $\epsilon$ as expected because the perimeter should
increase regardless of the sign of $\epsilon$. Since the area of the droplet
in contact with the plates remains constant, the relevant surface energy is
the surface energy along the perimeter,
\begin{equation}
\label{surf-nrg}
E_S = \sigma S \Delta,
\end{equation}
where $\sigma$ is the surface tension along the perimeter of the
droplet.  Note that we consider the case of $90^{\circ}$ contact angle
of the magnetic fluid-non magnetic fluid interface with the boundary
plates. Discussion of surface energy for other contact angles is 
in Ref.~\cite{banerjee}.

Minimizing the total energy $E_M+E_S$ with respect to $\epsilon$ gives
the elongation of a ferromagnetic droplet,
\begin{equation}
\label{ferro-elongation}
\epsilon_{\rm ferro}= {2 H_0^2 r_0^2 \over 3 \pi^2 \sigma \Delta}.
\end{equation}
The elongation is quadratic in the applied field since reversing the field
direction should not affect the result. At fixed thickness $\Delta$,
droplets with larger radius $r_0$
will elongate more. This is because magnetostatic energy
in~(\ref{mag-nrg4}) is proportional to $r_0^3$, whereas the surface energy is 
proportional to the area of the curved surface (of order $r_0 \Delta$). The 
elongation, however, is inversely proportional to $\Delta$. This is
because  the droplets with larger thickness find it easier 
to expel the applied field. The same applied field causes a relatively 
smaller distortion in the magnetization texture of a thicker droplet, and a 
smaller distortion of texture causes a smaller elongation. Figure 3 
shows the magnetization texture of a droplet as it elongates under 
applied field.

Determining the applied field at which the domain wall touches the
boundary and field penetration occurs is complicated by the dependence
of the reduced field $E_x$, defined in Eq.~(\ref{field}), on
elongation.  Basically, a given applied field ${\bf H}_0$ causes a greater
distortion of the magnetization texture in a droplet that is elongated
in the field direction than it would have caused in a circular
droplet. This is because the demagnetizing field becomes weak in elongated
droplets (that is why the droplet elongates, after all) so a larger charge
density is needed to achieve a demagnetizing field ${\bf H}_D=-{\bf H}_0$.
Correspondingly, a lower applied field is needed to achieve penetration
for an elongated droplet than for a circular droplet.

The reduced applied field in~(\ref{field}) can be written to first order
in $\epsilon$ as
\begin{equation}
E_x={H_0 r_0 \over 2 \pi \Delta M_s } (1 + {3 \epsilon \over 4}),
\end{equation}
with corrections of order $\epsilon^3$. Substituting the expression for 
elongation~(\ref{ferro-elongation}) into the above gives 
\begin{equation}
E_x={H_0 r_0 \over 2 \pi \Delta M_s} (1 + { H_0^2 r_0^2 \over 
2 \pi^2 \Delta \sigma}).
\end{equation}
For any finite $\sigma$, $E_x$ now depends nonlinearly on $H_0$.
$E_x$ is still a useful quantity as it measures the distortion in the
magnetization texture of the droplet, but its relation to applied
field is nontrivial.  It is therefore convenient to define a new
dimensionless field,
\begin{equation}
\label{dimless-field}
h^* \equiv {H_0 r_0 \over 2 \pi \Delta M_s},
\end{equation}
that equals $E_x$ when the droplet is circular.
We also define a dimensionless surface tension 
\begin{equation}
\sigma^*={\sigma \over 2 M_s^2 \Delta},
\end{equation}
so that the reduced field can be written as
\begin{equation}
\label{dimless}
E_x=h^* (1+{{h^*}^2 \over \sigma^*}).
\end{equation}
Our calculations hold in the limit of small droplet distortion,
${h^*}^2/\sigma^* << 1$, and non-penetration of the field, $E_x < 1$.
Consider the applied field at which field penetration occurs. Solving
Eq.~(\ref{dimless}) for the critical applied field $h_c^*$ for
which $E_x=1$ yields $h_c^* \approx 1-1/\sigma^*$ in the limit of
nearly undeformable droplets, $\sigma^* >> 1$.

Figure~\ref{f3} illustrates the simultaneous evolution of droplet
shape and texture. Three stages are shown, at $E_x=0, 0.5$ and $1$.
Parameters are chosen so that $\sigma^*=1$. As expected, in contrast
to the values of $E_x$, the actual applied field values $h^*$ become
more closely spaced as the elongation grows, and $h_c^* =0.68< 1$.

\section{discussion}
\label{discussion}

\subsection{Contrast with paramagnetic thin films}
\label{contrast}
Expression~(\ref{ferro-elongation}) for elongation of ferromagnetic droplets 
contrasts with the elongation of paramagnetic droplets in the same thin film 
geometry. For a paramagnetic thin film with susceptibility $\chi$, the 
elongation has been previously calculated~\cite{banerjee},
\begin{equation}
\label{para}
\epsilon_{\rm para}={\chi^2 H_0^2 \Delta \over \sigma} \ln {C 
r_0 \over \Delta},
\end{equation}
where $C$ is a constant.

To understand  this 
contrasting behavior we provide the following simple argument on dimensional
grounds. The elongation is a ratio of the magnetostatic and surface energies. 
Let $Q$ be the net charge on the portion of the droplet with $x>0$. For 
$x<0$ the net charge is $-Q$. In both the ferromagnetic and paramagnetic 
cases, the charges on average are separated by distances of order $r_0$, 
and therefore the magnetostatic energy goes as $Q^2/r_0$. Dividing this by 
the surface energy (of order $r_0 \Delta \sigma$) gives the elongation
\begin{equation}
\label{para2}
\epsilon \sim {Q^2 \over r_0^2 \sigma \Delta},
\end{equation}
where ``$\sim$'' indicates proportionality.
In the case of a ferromagnetic droplet $Q \sim H_0 r_0^2$ because the
charge density $\rho_M$ is distributed in the bulk of the droplet
(volume $\sim r_0^2 \Delta$) but is inversely proportional to $\Delta$.
This explains why the elongation of a ferromagnetic droplet increases
rapidly with $r_0$ and decreases with $\Delta$ in Eq.~(\ref{ferro-elongation}). 
In the case of a paramagnetic droplet, the charges are distributed mainly along the
curved surface of the droplet, so the total charge $Q$ is therefore
proportional to $\chi H_0 r_0 \Delta$. Our simple dimensional argument gives
$\epsilon \sim \chi^2 H_0^2 \Delta/\sigma$, which reproduces Eq.~(\ref{para}) except
for the dimensionless $\ln (r_0/\Delta)$, and explains why the
elongation in the paramagnetic case depends only weakly on $r_0$ and
increases strongly with $\Delta$.

Although theory and computer simulations predict the existence of a 
ferromagnetic liquid state there is still not much experimental evidence 
for it. The contrasting behavior of ferromagnetic thin films and
paramagnetic thin films under applied fields could serve as a useful test 
to detect such a state. This contrasting behavior can be neatly captured
in a picture if one plots $\epsilon / \Delta$ vs the inverse aspect ratio,
$2r_0 /\Delta$ at a fixed field value. Figure 4 shows such a plot for 
experiments on ferromagnetic and paramagnetic droplets of a hypothetical 
fluid at  $H_0=20$ G. The fluid has a susceptibility $\chi = 5$ 
in the paramagnetic phase, surface tension $\sigma= 60$ dynes/cm and
$M_s=500$ G, typical of ferrofluids.  We take the droplet dimensions $2r_0=120~ \mu$m 
and $\Delta=1.2~ \mu$m so that  $p=2r_0/\Delta=10^{-2}$ and $\sigma^*=1$.     
The value $H_0=20$ G corresponds to $h^*=0.16$ for the droplet in the ferromagnetic phase. 
Recall that for $\sigma^*=1$ field penetration occurs at $h^*=0.68$.

\subsection{Symmetry breaking of shape under exchange interaction}
\label{symbreak}

Despite the asymmetric magnetization texture, the droplet shape in Fig. 3 has 
reflection symmetry about the field
direction. Interactions of the magnetic poles with the applied field and with each
other create the forces that deform the droplet.  Since, for a symmetric 
shape the induced charge density is symmetric, the droplet deformation is symmetric. 
The deformed magnetization texture is asymmetric because the direction of rotation of
magnetization lines (counterclockwise in Fig. 2 and 3) breaks the left-right symmetry.
If the exchange interaction due to spatial variation of magnetization were to be
included, the droplet shape would become asymmetric to lower its
exchange energy.

For an isotropic medium, the exchange energy density can be written as~\cite{landau}
\begin{equation}
\label{exchange2}
U_{EX}= {1 \over 2} \alpha  {\partial M_k \over \partial x_i} {\partial M_k 
\over \partial x_i},
\end{equation}
where $\alpha$ is the exchange constant and the summation convention is employed. The 
magnetization texture for a droplet under applied field breaks the 
left-right symmetry (see Figs. 2 and 3). The
breaking of  the symmetry is most evident at the domain wall where $U_{EX}$ 
is singular. The exchange interaction  creates forces that break the 
left-right symmetry of the droplet shape.

The main contribution to the exchange energy of the droplet comes from near the domain 
wall, where the magnetization has an apparent discontinuity of order $M_s$. The domain 
wall actually spreads to a finite width $w$ to lower its exchange energy at the cost
of acquiring a demagnetizing energy.  Since the magnetization changes by an amount of 
order $M_s$ over
a distance $w$, the domain wall exchange energy density is roughly $\alpha M_s^2 /w^2$. 
For a Bloch domain wall~\cite{domainwall} with $w<< \Delta$ the demagnetizing energy 
density is of order $M_s^2 w/\Delta$. 
Minimizing the total energy  per unit domain wall length with 
respect to the width gives $w$ of order $(\alpha \Delta)^{1/3}$ and energy density of
order $\alpha^{1/3} M_s^2 /\Delta^{2/3}$. 

For $E_x << 1$, symmetry  arguments and numerical calculations suggest that domain 
wall arc-length varies as $E_x^2 r_0$. The total energy of the domain wall can 
be estimated by multiplying the domain wall energy density with the domain 
wall arc-length ($E_x^2 r_0$) and cross section ($w \Delta$) to get
\begin{equation}
E_{wall} \sim \alpha^{2/3} M_s^2 E_x^2 \Delta^{2/3} r_0=\alpha^{2/3} H_0^2 {r_0^3 \over 
\Delta^{4/3}}.
\end{equation}
The ratio of the domain wall energy to 
the magnetostatic energy~(\ref{mag-nrg4}) goes like $(\alpha / \Delta^2)^{2/3}$. 
Therefore, the relative importance of exchange effects diminishes for large thickness 
$\Delta$.

To understand how exchange interaction breaks the symmetry of the shape, we analyze
the exchange forces on a circular disk in an applied field (Fig. 2). We write the  
shape of the droplet as a perturbed circle breaking the left-right symmetry, 
\begin{equation}
r(\theta) = r_0 { (}1-{\psi^2 \over 2}  + \psi \sin 3 \theta { )}.
\end{equation}
Here $\theta$ is the angle measured counterclockwise from the direction of the 
applied field in the plane of the droplet.  The 
$\sin 3\theta$ term is the lowest harmonic in $\theta$ which breaks the left-right 
symmetry of the shape. The perturbation keeps the area  constant to
ensure that the volume of the droplet does not change. Assuming analytic variation 
with respect to the perturbation and the applied field, the domain wall energy takes 
the following form for small $\psi$ and $E_x$,
\begin{equation}
\label{ex-nrg}
E_{wall} = \alpha^{2/3} M_s^2 E_x^2 {r_0 \Delta^{2/3} }(A + B \psi E_x),
\end{equation}
where $A,B$ are dimensionless constants.  The order $\psi$ term is multiplied by 
$E_x$ because under simultaneous reversals 
of $\psi$ and $E_x$ the domain wall energy remains the same.

The surface energy of the droplet varies quadratically with $\psi$,
\begin{equation}
E_s= 2 \pi r_0 \Delta \sigma (1 + 2 \psi^2).
\end{equation}
The magnetostatic energy~(\ref{mag-nrg4}) of the droplet  has an order $\psi^2$ 
correction that can be neglected because it is higher order in the applied field. 
Minimizing $E_S+E_{wall}$ with respect to $\psi$ gives
\begin{equation}
\psi = -{B \alpha^{2/3} M_s^2 E_x^3 r_0 \over 8 \pi \Delta^{1/3} \sigma}.
\end{equation}
We conclude that for weak applied fields the symmetry breaking 
in shape will be much smaller than the elongation of the droplet which goes like 
$E_x^2$. The direction of symmetry breaking (sign of $\psi$) remains 
undetermined, since it depends on the sign of $B$. We expect the direction of 
symmetry breaking to shorten the length of the domain wall.

\section{conclusion}
\label{conclusion}

We calculate the magnetization texture and elongation of a ferromagnetic
thin film under weak applied fields. The point vortex favored at
zero field elongates into a curved domain wall in applied field. We
find that the droplet elongation is proportional to the square of the radius
of the undeformed droplet and inversely proportional to its thickness.
When domain wall energies are considered, the ferromagnetic droplet may
 break reflection symmetry. These results contrast with
that of paramagnetic thin films. We propose this different behavior
under applied field as one possible way to detect the ferromagnetic
state in a liquid.

\acknowledgements

We acknowledge useful discussions and communications with J. Reske, S. Dietrich, 
R. B. Griffiths, A. A. Thiele and L. Berger. This work was supported in part by NSF grant 
DMR-9732567.

\begin{figure}[tb]
\epsfxsize=6in \epsfbox{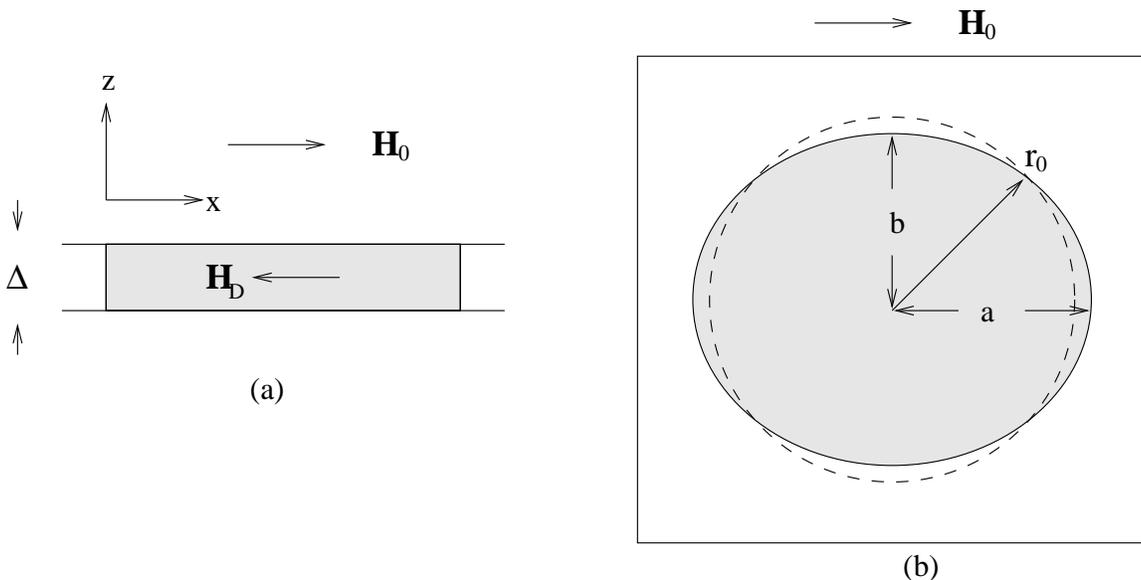}
\caption{(a) A side view of a ferromagnetic liquid droplet confined between 
two parallel plates.
(b) A top view of the droplet elongating under applied field. The 
dashed line shows the undeformed droplet.}
\label{f1}
\end{figure}

\begin{figure}[tb]
\psfig{figure=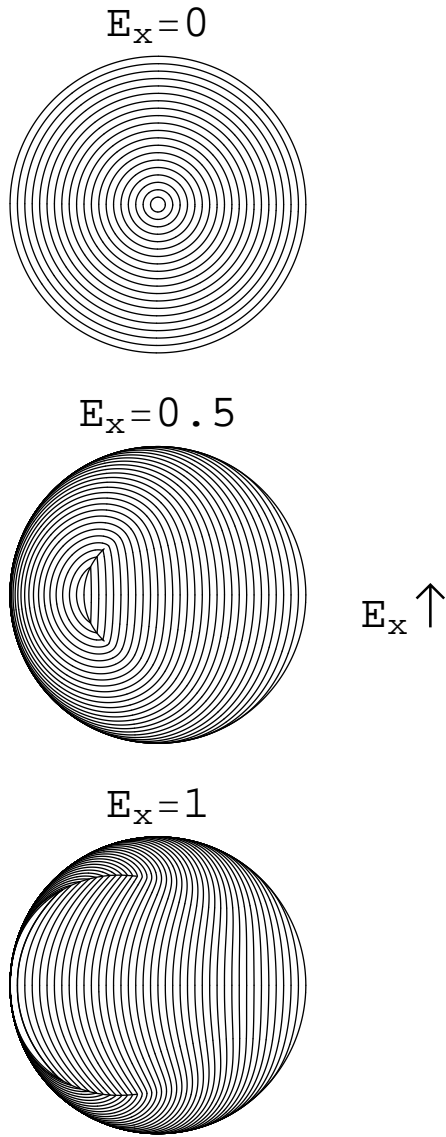,height=8.0in}
\caption{The magnetization texture of a solid circular ferromagnetic thin film under 
applied field~[22]. $E_x$ is the reduced applied field defined in Eq.~(\ref{field}). 
The magnetization texture in zero field is going counterclockwise.}
\label{f2}
\end{figure}
\begin{figure}[tb]
\psfig{figure=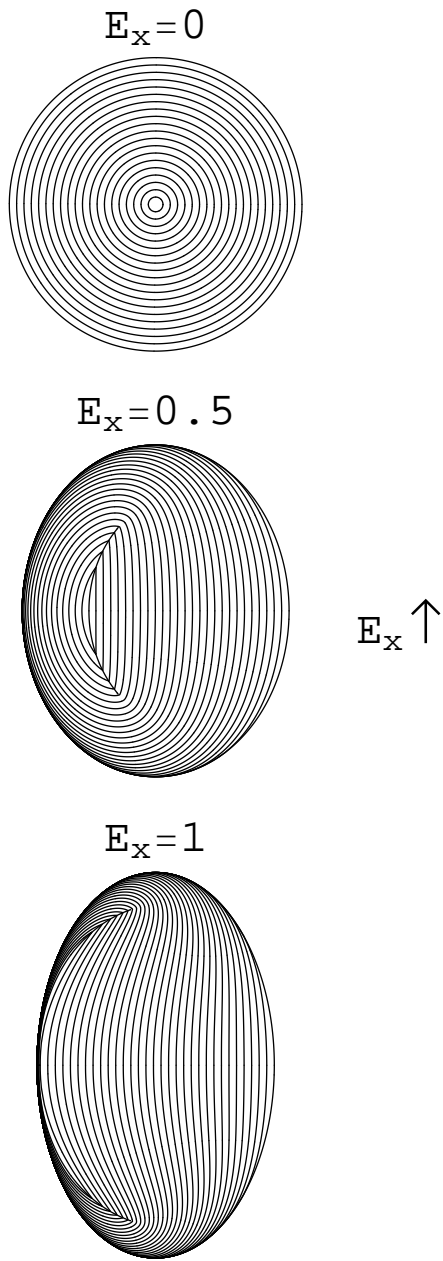,height=8.0in}
\caption{The magnetization texture of a liquid ferromagnetic thin film elongating 
under applied field. 
The magnetization texture in zero field is going counterclockwise.
Values for dimensionless field $h^*$ (see
Eq.~(\ref{dimless-field})) are 0, 0.42 and 0.68.}
\label{f3}
\end{figure}

\begin{figure}[tb]
\psfig{figure=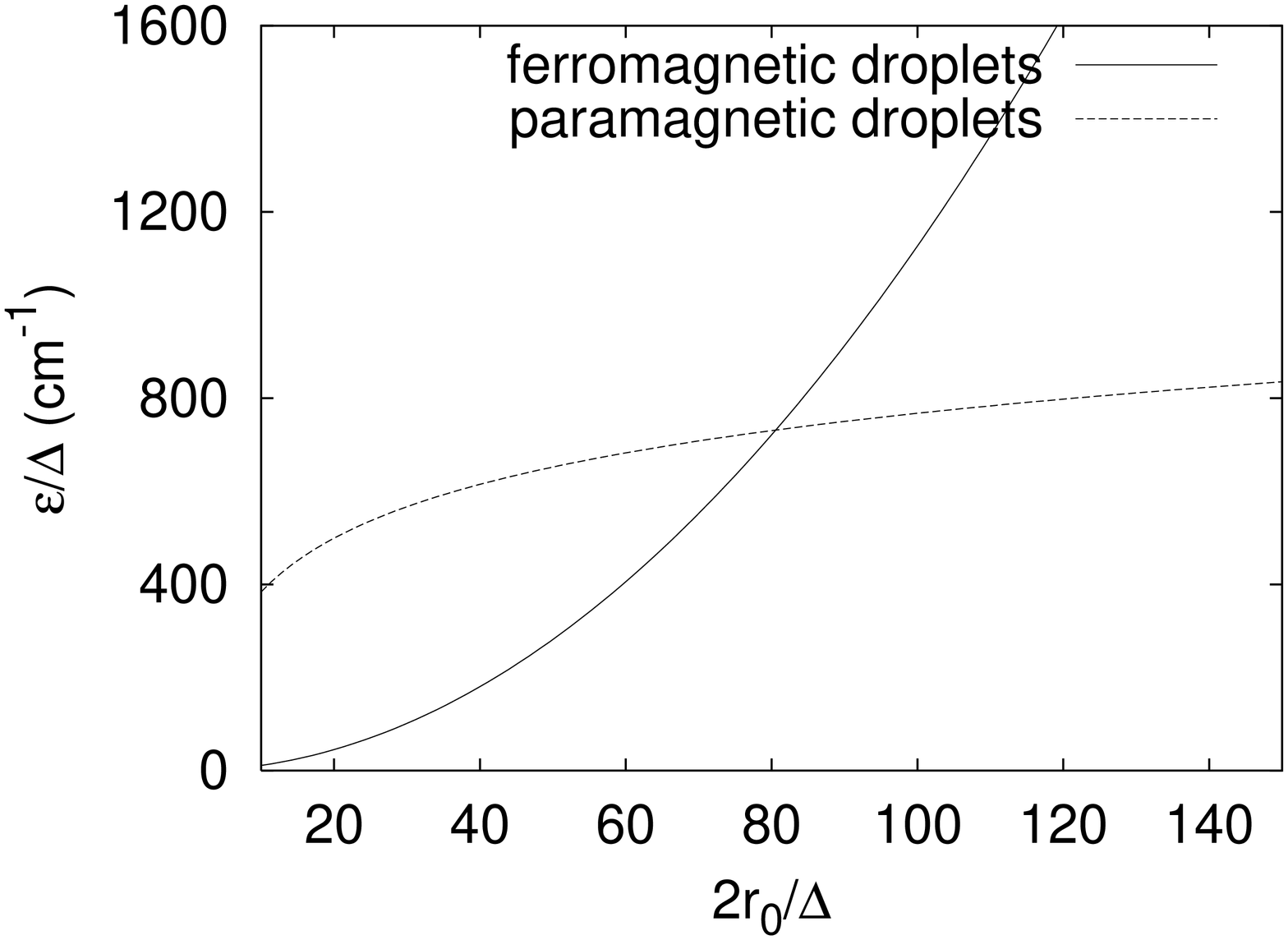,angle=-90,height=4.0in,width=6.0in}
\vspace{0.3cm}
\caption{Plots of $\epsilon /\Delta$ vs $2r_0/\Delta$ for ferromagnetic and
paramagnetic droplets for a fixed value of $H_0=20$ G.}
\label{f4}
\end{figure}


\begin{references}

\bibitem{sano} K. Sano and M. Doi, J. Phys. Soc. Jpn. {\bf 52}, 2810 (1983).

\bibitem{tsebers} A. O. Tsebers, Magnetohydrodynamics {\bf 2}, 42 (1982). 

\bibitem{widom} H. Zhang and M. Widom, J. Magn. Magn. Mater. {\bf 122}, 119 
(1993); Phys. Rev. E {\bf 49}, R3951 (1994).

\bibitem{dietrich} B. Groh and S. Dietrich, Phys. Rev. Lett. {\bf 72} 2422 
(1994); {\bf 74}, 2617 (1995); Phys. Rev. E {\bf 50}, 3814 (1994).

\bibitem{groh} B. Groh and S. Dietrich, Phys. Rev. E {\bf 57}, 4535 (1998).

\bibitem{patey} D. Wei and G. N. Patey, Phys. Rev. Lett. {\bf 68}, 2043 
(1992); Phys. Rev. A {\bf 46}, 7783 (1992).

\bibitem{weis1} J. J. Weis, D. Levesque, and G. J. Zarragoicoechea, Phys. Rev. 
Lett. {\bf 69}, 913 (1992).

\bibitem{weis2} J. J. Weis and  D. Levesque, Phys. Rev. E {\bf 48}, 3728 (1993).

\bibitem{stevens} M. J. Stevens and G. S. Grest, Phys. Rev. E {\bf 51}, 5962
(1995); Phys. Rev. E {\bf 51}, 5976 (1995).

\bibitem{rr} R. E. Rosensweig, {\it Ferrohydrodynamics} (Cambridge University Press,
Cambridge, England, 1985).

\bibitem{luo} J. Zhang, C. Boyd, and W. Luo, Phys. Rev. Lett. {\bf 77}, 390 (1996).

\bibitem{ederbeck} D. Ederbeck and H. Ahlers, J. Magn. Magn. Mater. {\bf 192}, 148 
(1999).

\bibitem{luo2} W. Luo, Nuovo Cimento D (Italy) {\bf 16}, 1199 (1994).

\bibitem{platzek} J. Reske, D. M. Herlach, F. Keuser, K. Maier, and D. Platzek,
Phys. Rev. Lett. {\bf 75}, 737 (1995); T. Albrecht, C. Buhrer, M. Fahnle, K. Maier, 
D. Platzek, and J. Reske, Appl. Phys. A {\bf 65},  215 (1997).

\bibitem{liquid-crystal} R. B. Meyer, L. Liebert, L. Strzelecki, P. Keller, 
J. de Phys. (France) Lett.  {\bf 30}, 69 (1975).

\bibitem{helium} N. D. Mermin and D. M. Lee, Scientific American {\bf 235}, 56 
(1976).

\bibitem{degennes} P. G. de Gennes and P. A. Pincus, Solid State Comm. {\bf 7},
339 (1969).

\bibitem{wfb}  W. F. Brown, {\it Magnetostatic Principles in Ferromagnetism}
(North-Holland, Amsterdam, 1962).

\bibitem{mermin} N. D. Mermin, Rev. Mod. Phys. {\bf 51}, 591 (1979).

\bibitem{hubert} A. Hubert, J. de. Phys. (France) {\bf C8}, 1859 (1988).

\bibitem{vandenberg} H. A. M. van den Berg, J. Appl. Phys. {\bf 60}, 1104 (1986).

\bibitem{suhl} P. Bryant and H. Suhl, Appl. Phys. Lett. {\bf 54}, 2224 (1989).

\bibitem{banerjee} S. Banerjee, M. Fasnacht, S. Garoff, and M. Widom, Phys. Rev. E. 
{\bf 60}, 4272 (1999).

\bibitem{landau} L.D. Landau and E.M. Lifshitz, {\it Electrodynamics of Continous 
Media} (Addison Wesley, 1960).

\bibitem{domainwall} See  S. Middelhoek, J. Appl. Phys. {\bf 34}, 1054 (1963) for a 
review article on domain walls in ferromagnetic thin films.

\end{references}
\end{document}